\documentclass{ICEAA-IEEE_APWC}
\pdfoutput=1

\usepackage[numbers,sort]{natbib}
\usepackage{amsmath}
\usepackage{booktabs}
\usepackage{color}
\usepackage{aastex_hack}

\makeatletter
\let\@fnsymbol\@arabic
\makeatother

\title{Testing a modified {ASKAP} Mark II phased array feed on the 64 m {P}arkes radio telescope}

\author{%
A. P. Chippendale\thanks{CSIRO Astronomy and Space Science, PO Box 76, Epping 1710, Australia, e-mail:
\texttt{Aaron.Chippendale@csiro.au}.} \and R. J. Beresford$^1$ \and X. Deng$^{1,}$\thanks{Max-Planck-Institut f{\"u}r Radioastronomie (MPIfR), Auf dem H{\"u}gel 69, D-53121 Bonn, Germany.} \and M. Leach$^1$ \and J. E. Reynolds$^1$ \and M. Kramer$^2$ \and T. Tzioumis$^1$ } 

\begin{document}
\maketitle

\begin{abstract}
We present the first installation and characterization of a phased array feed (PAF) on the 64 m Parkes radio telescope.  The combined system operates best between 0.8~GHz and 1.74~GHz where the beamformed noise temperature is between 45~K and 60~K, the aperture efficiency ranges from 70\% to 80\%, and the effective field of view is 1.4~deg$^2$ at 1310~MHz.  After a 6-month trial observing program at Parkes, the PAF will be installed on the 100~m antenna at Effelsberg. This is the first time a PAF has been installed on a large single-antenna radio telescope and made available to astronomers.
\end{abstract}

\section{INTRODUCTION}
Installing a phased array feed (PAF) at the focus of a concentrator, together with digital beamforming, enhances the antenna's field of view, survey speed, and operational flexibility. This is achieved by forming and processing beams covering many adjacent directions on the sky at the same time. 

The Max Planck Institute for Radio Astronomy (MPIfR) is collaborating with CSIRO to deploy a PAF on the Parkes 64~m radio telescope and later on the Effelsberg 100~m telescope.   MPIfR and CSIRO will initially use the PAF to look for fast radio bursts (FRBs).  These are unexplained radio emissions that last only of order milliseconds but appear to come from the distant universe \cite{Thornton2013}.  The PAF monitors more of the sky at any instant and so increases the chances of discovery and localisation.

CSIRO designed the PAF for the Australian SKA Pathfinder telescope (ASKAP) \cite{DeBoer2009} to demonstrate fast astronomical surveys with a wide field of view for the Square Kilometre Array (SKA) over 0.7~GHz to 1.8~GHz.  The SKA is an international project to build the world's largest radio telescope, with one square kilometre of collecting area. 

The ASKAP PAF was designed for the remote Australian SKA site at the Murchison Radio-astronomy Observatory (MRO) within a legislatively protected radio-quiet zone \cite{Wilson2013}.  To operate at Parkes and Effelsberg, the PAF was fitted with narrower sampling filters that accept the 1.2--1.75~GHz band and reject interference from mobile phone and digital television services at lower frequencies that are far more prevalent at Parkes and Effelsberg.


\section{THE MPIfR PAF}
\begin{figure}[t]
\centerline{\includegraphics[width=5.5cm]{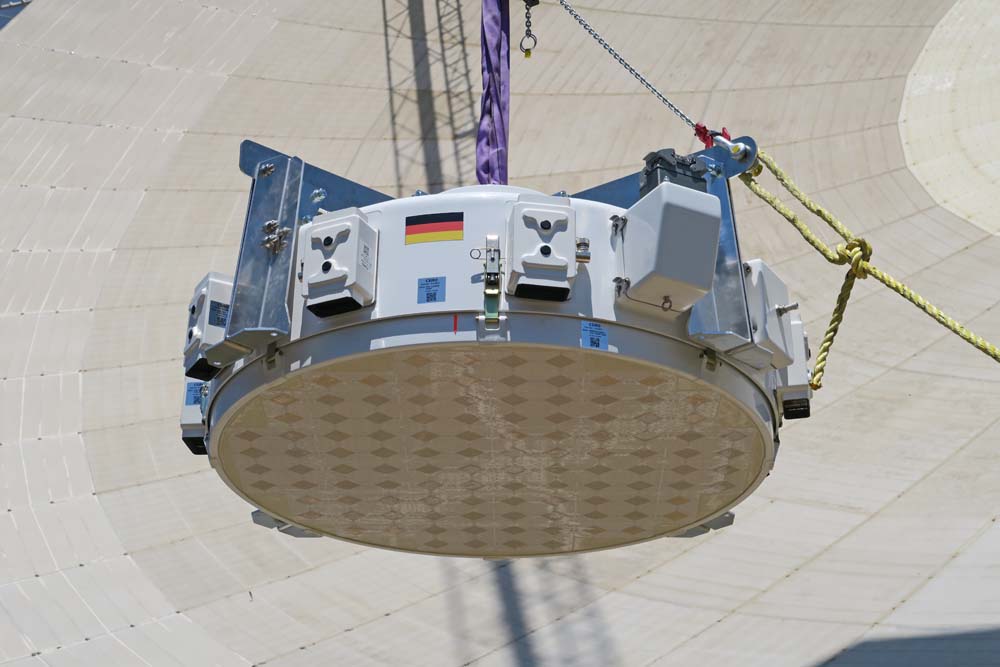}} \caption{\small Lifting the Mk.~II ASKAP PAF, modified for MPIfR, to the focus of the Parkes 64 m radio telescope (credit: J. Sarkissian). \label{fig:paflift}}
\end{figure}
The MPIfR PAF is a modified Mk.~II ASKAP PAF \cite{Hampson2012} based on a connected-element ``chequerboard'' array \cite{Hay2008} that is dual-polarized, low-profile, and inherently wide-band.   Figure \ref{fig:paflift} shows the MPIfR PAF being lifted to the focus of the Parkes 64 m radio telescope on 10 February 2016.

To operate an ASKAP PAF in the populous surrounds of Parkes and Effelsberg required modified band preselection to withstand significant terrestrial radio-frequency interference (RFI).  The ASKAP and MPIfR filter definitions are compared in Table \ref{tab:filters}.  Reduced MPIfR bandpasses in bands 2 and 3 keep the signal chain linear by rejecting broadband telecom in 694--820~MHz, 850--960~MHz and 1805--1880~MHz, and aeronautical ADS-B/DME at 960--1164~MHz.

\begin{table}[b]
\begin{center}
\footnotesize
\begin{tabular}{ccc}
\toprule
  Band & ASKAP & MPIfR \\ 
       & (MHz) & (MHz) \\ \midrule
  1 & \phantom{1}700--1200 & \phantom{1}700--1200 \\
  2 & \phantom{1}840--1440 & 1200--1480 \\
  3 & 1400--1800 & 1340--1740 \\
  4 & \phantom{1}600--700\phantom{0}  & \phantom{1}600--700\phantom{0}  \\
  \bottomrule
\end{tabular}
\end{center}
\caption{ASKAP and MPIfR PAF filters.}\label{tab:filters}
\end{table}

The Mk.~II PAF system has band-select filters at both ends of all 188 radio frequency over fibre (RFoF) links between the PAF and the digital receiver.  In the MPIfR PAF system, only the band-select filters in the PAF have been modified.  The filters in the digital receivers are ASKAP units.  

The MPIfR PAF requires 60~dB of RF gain to achieve lower than 2~K back-end contribution to system noise and position the drive level to the 12-bit analog-to-digital converter (ADC) inputs at 40 quantisation levels. The 1~dB compression point referred to the array element feed lines is typically better than -60~dBm with 25~dB dynamic headroom to cope with unwanted RFI signals, the main limitation coming from the spurious free dynamic range of the RFoF subsystem, which is  110~$\text{dB} \, \text{Hz}^{-2/3}$.  

\begin{figure*}[htpb]
\centerline{\includegraphics[width=\textwidth, trim=12mm 0mm 12mm 0mm, clip]{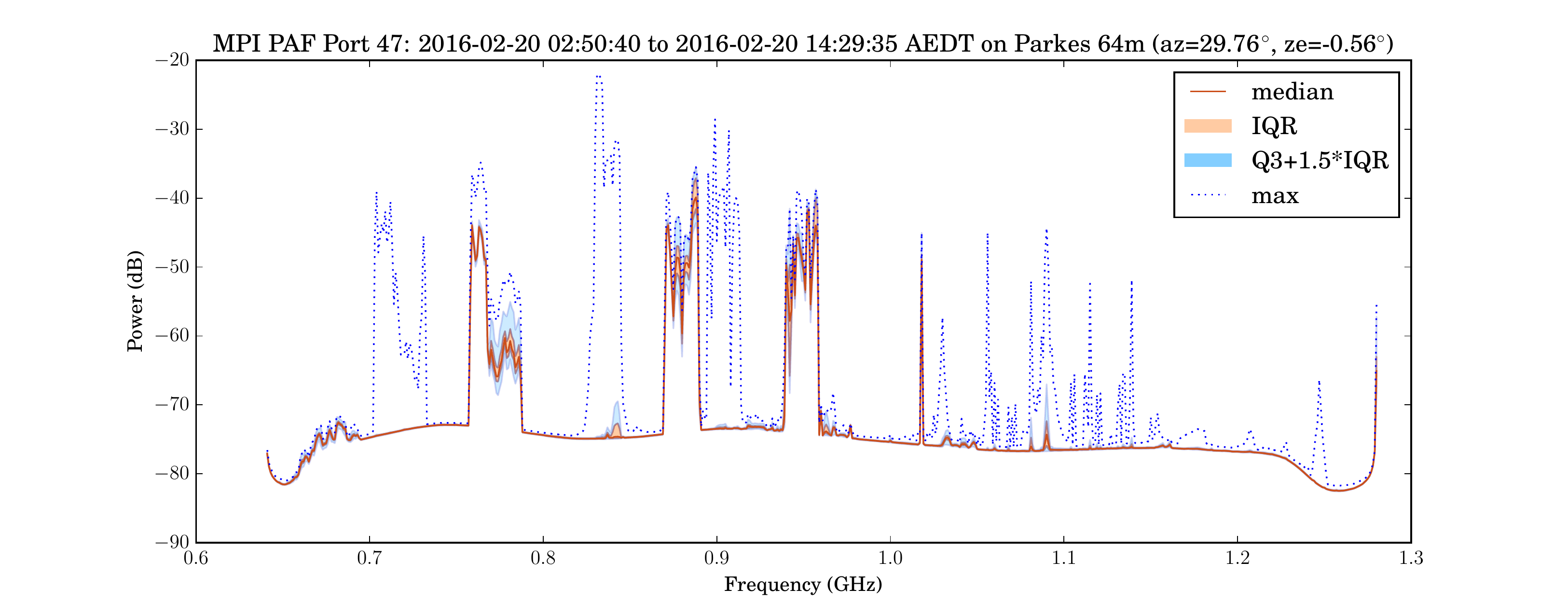}} \caption{\small Band 1 spectrum statistics at Parkes with 1~MHz resolution. \label{fig:band1cfb}}
\end{figure*}

\begin{figure*}[htpb]
\centerline{\includegraphics[width=\textwidth, trim=12mm 0mm 12mm 0mm, clip]{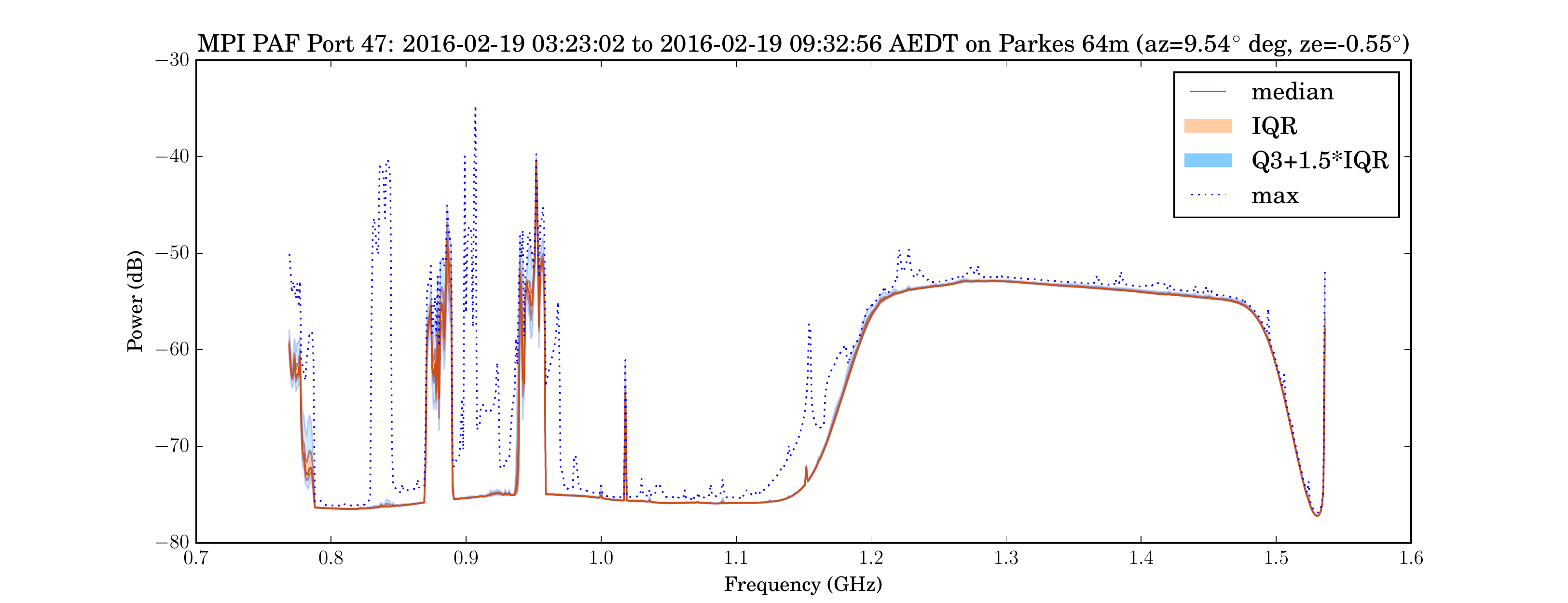}} \caption{\small Band 2 spectrum statistics at Parkes with 1~MHz resolution. \label{fig:band2cfb}}
\end{figure*}

\begin{figure*}[htpb]
\centerline{\includegraphics[width=\textwidth, trim=12mm 0mm 12mm 0mm, clip]{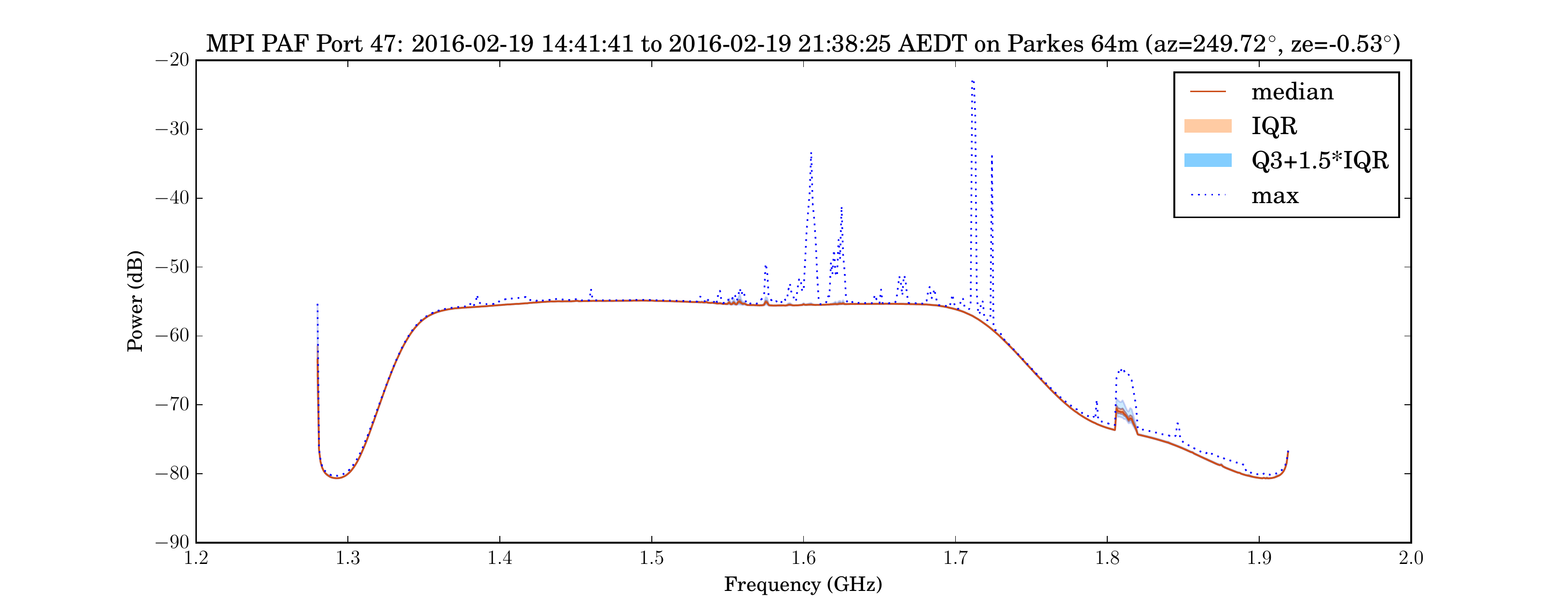}} \caption{\small Band 3 spectrum statistics at Parkes with 1~MHz resolution. \label{fig:band3cfb}}
\end{figure*}

Figures \ref{fig:band1cfb}--\ref{fig:band3cfb} show statistics of the 1~MHz resolution spectrum of a single port near the centre of the MPIfR PAF.  These measurements were taken with the PAF installed on the Parkes 64 m antenna while it was stowed and pointing near the zenith.

Rejection of RFI by the modified filters in bands 2 and 3 allows the MPIfR PAF to work at Parkes with the same attenuator settings and ADC input levels used for ASKAP at the radio-quiet MRO.  In band 1, the MPIfR PAF RF signals require at least 10 dB more attenuation than at the MRO.  

\section{INSTALLATION AT PARKES}

The PAF was installed at the focus of the Parkes 64 m antenna which is a paraboloidal reflector with a focal ratio $F/D$ of 0.41.  The phase centre of the PAF, the antenna-side surface of the PAF groundplane, was positioned to be 24.9$\pm$5~mm above the optical focus of the 64~m reflector when the telescope is tipped to 45$^\circ$ elevation.  The PAF was installed so that, if the dish were hypothetically tipped to the horizon, the principal polarisation planes of the PAF would be 7.5$^\circ$ clockwise from both the vertical and the horizontal as viewed from the reflector's vertex looking towards the PAF.  

To aid calibration we have installed a radiator at the vertex of the Parkes reflector to transmit broadband noise into the PAF.  This is a log-periodic dipole array antenna (Aaronia HyperLOG 7025) with linear polarisation that can be manually rotated, but is nominally left at 45$^\circ$ to the polarisation planes of the PAF elements.  PAF port gains may be estimated via correlation of the the PAF RF signals with a direct copy of the radiated calibration noise \cite{Chippendale2016}.  Additional fibre was placed in the link carrying the reference copy of the calibration noise so it reaches the digital receiver with the same delay as the noise that is radiated into the PAF. 

Initial measurements suggest that calibration via the vertex source will be more complicated on the Parkes 64 m than the promising first attempts on an ASKAP 12 m antenna in \cite{Chippendale2016}.  This is due to multipath interference caused by the reflection of the calibration noise from the large floor of the focus cabin at Parkes and then the concentration of this reflection back into the PAF by the paraboloidal reflector.  Moving the calibration source away from the vertex may mitigate this effect.  


\section{PERFORMANCE MEASUREMENTS}

\begin{figure}[tb]
\centerline{\includegraphics[width=\columnwidth, trim=0 2mm 0 12mm, clip]{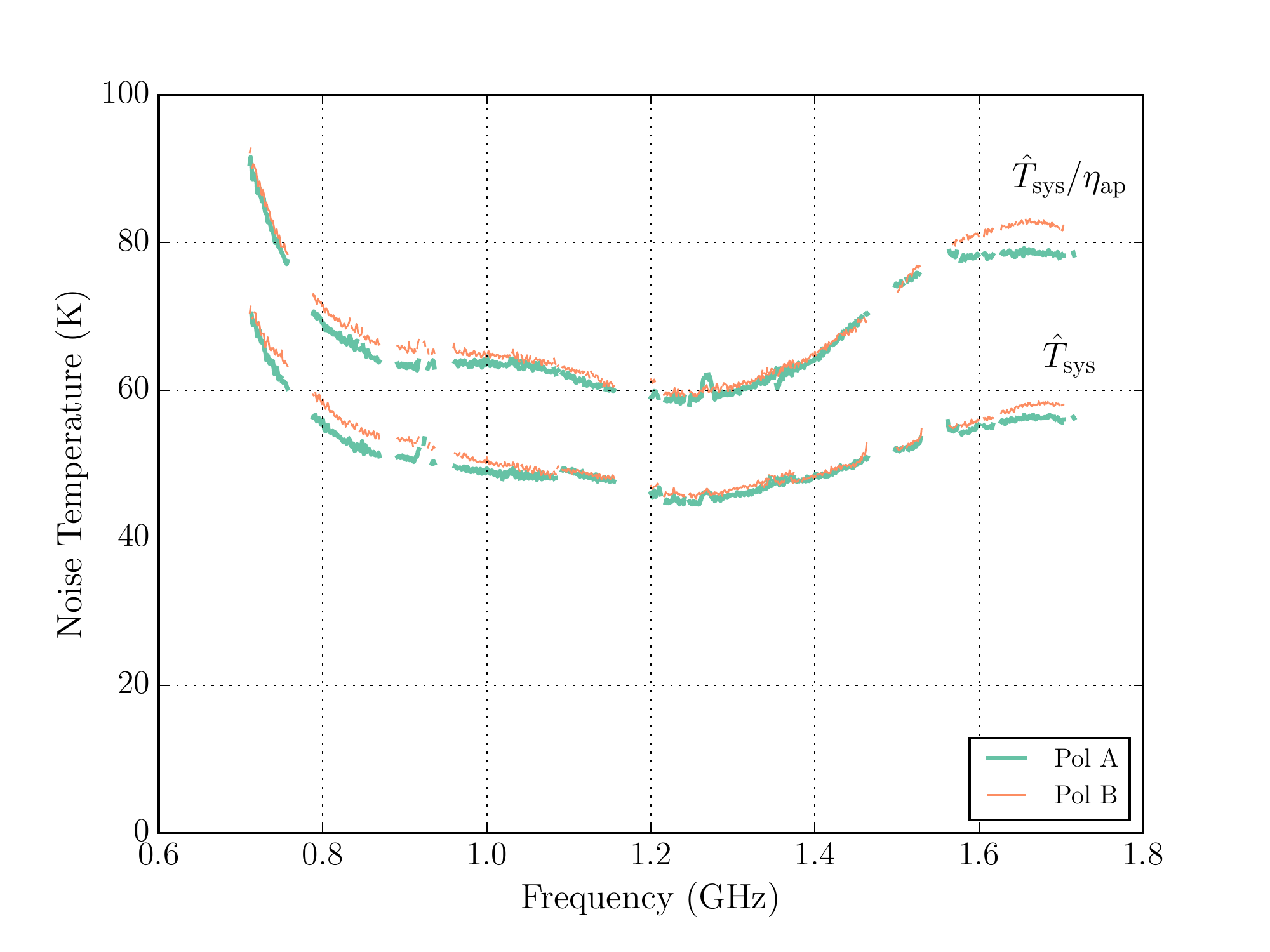}} \caption{\small Noise temperature of the MPIfR PAF on the Parkes 64~m antenna.  Boresight maxSNR beam. \label{fig:tsys}}
\end{figure}

\begin{figure}[tb]
\centerline{\includegraphics[width=\columnwidth, trim=0 2mm 0 12mm, clip]{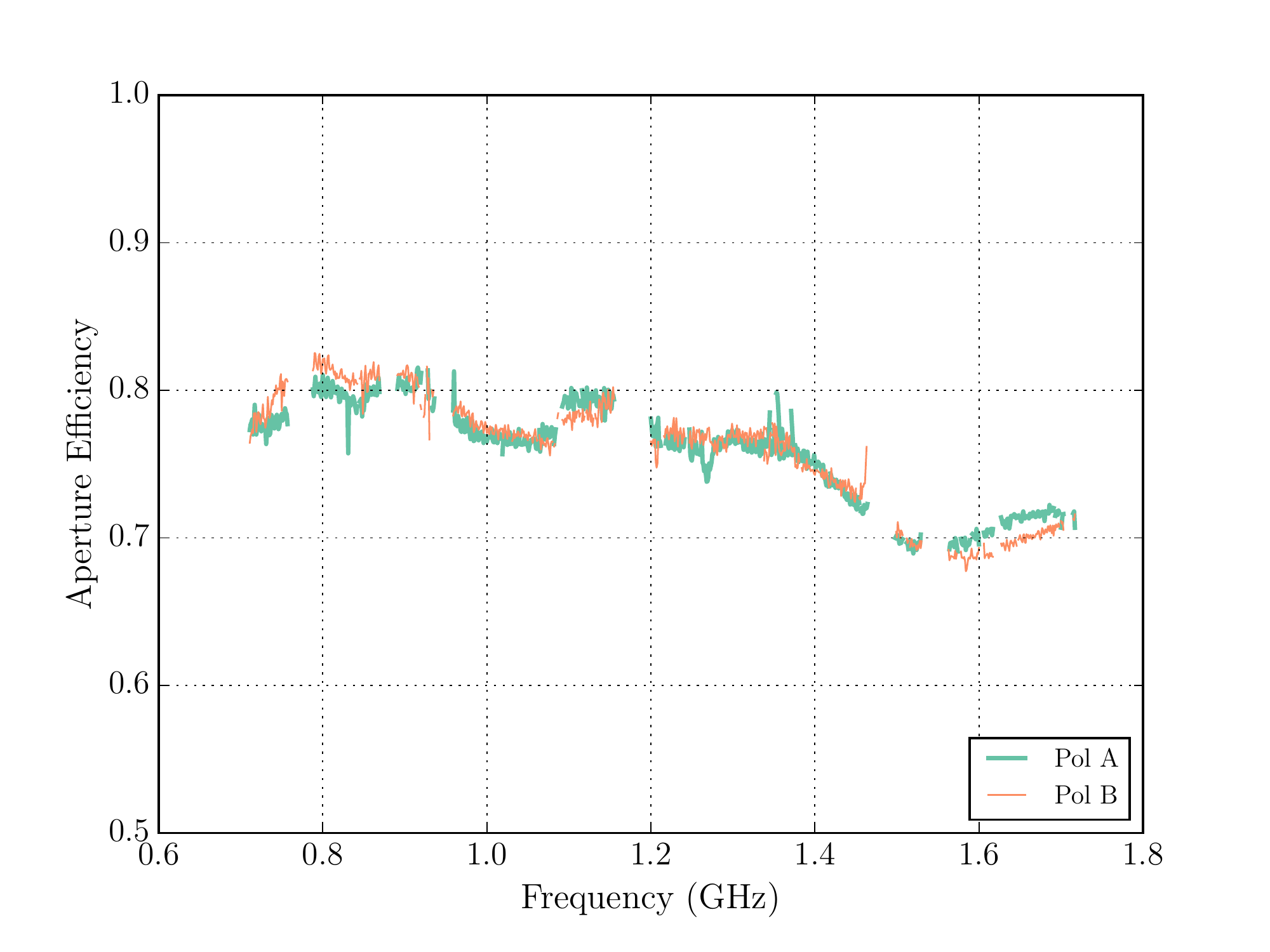}} \caption{\small Aperture efficiency $\eta_\text{ap}$ of the MPIfR PAF on the Parkes 64~m antenna. Boresight maxSNR beam. \label{fig:eta}}
\end{figure}

Figure \ref{fig:tsys} shows noise temperatures measured for a boresight beam with maximum signal-to-noise ratio (maxSNR) weights calculated as in \cite{Chippendale2015}.  The lower curve is the beam equivalent system noise temperature $\hat{T}_\text{sys}$ referred to the sky.  This is measured via a Y-factor measurement on microwave absorber at ambient temperature, temporarily inserted in front of the PAF while it was at the focus of the Parkes 64 m antenna.  The upper curve is the system-temperature-over-efficiency $\hat{T}_\text{sys}/\eta_\text{ap}$ measured via a Y-factor measurement on the radio source Virgo~A with known spectral-flux density \cite{Ott1994}.  We used Vir~A both to calculate the beam weights and as the flux standard.  Beamforming and flux calibration were performed as per \cite{Chippendale2015} and the hot-load Y-factor measurements follow \cite{Chippendale2015a} but use the same beam weights and unobstructed sky reference measurements as the Vir~A Y-factor measurements.   

Figure \ref{fig:eta} shows an estimate of the aperture efficiency $\eta_\text{ap}$ formed by dividing $\hat{T}_\text{sys}$ by $\hat{T}_\text{sys}/\eta_\text{ap}$.  All sky measurements were made between elevations of 35.4$^\circ$ and 48.0$^\circ$.  Gaps in frequency coverage are due to the removal of data affected by RFI.  

\begin{figure}[tb]
\centerline{\includegraphics[width=\columnwidth, trim=15mm 0mm 0mm 10mm, clip]{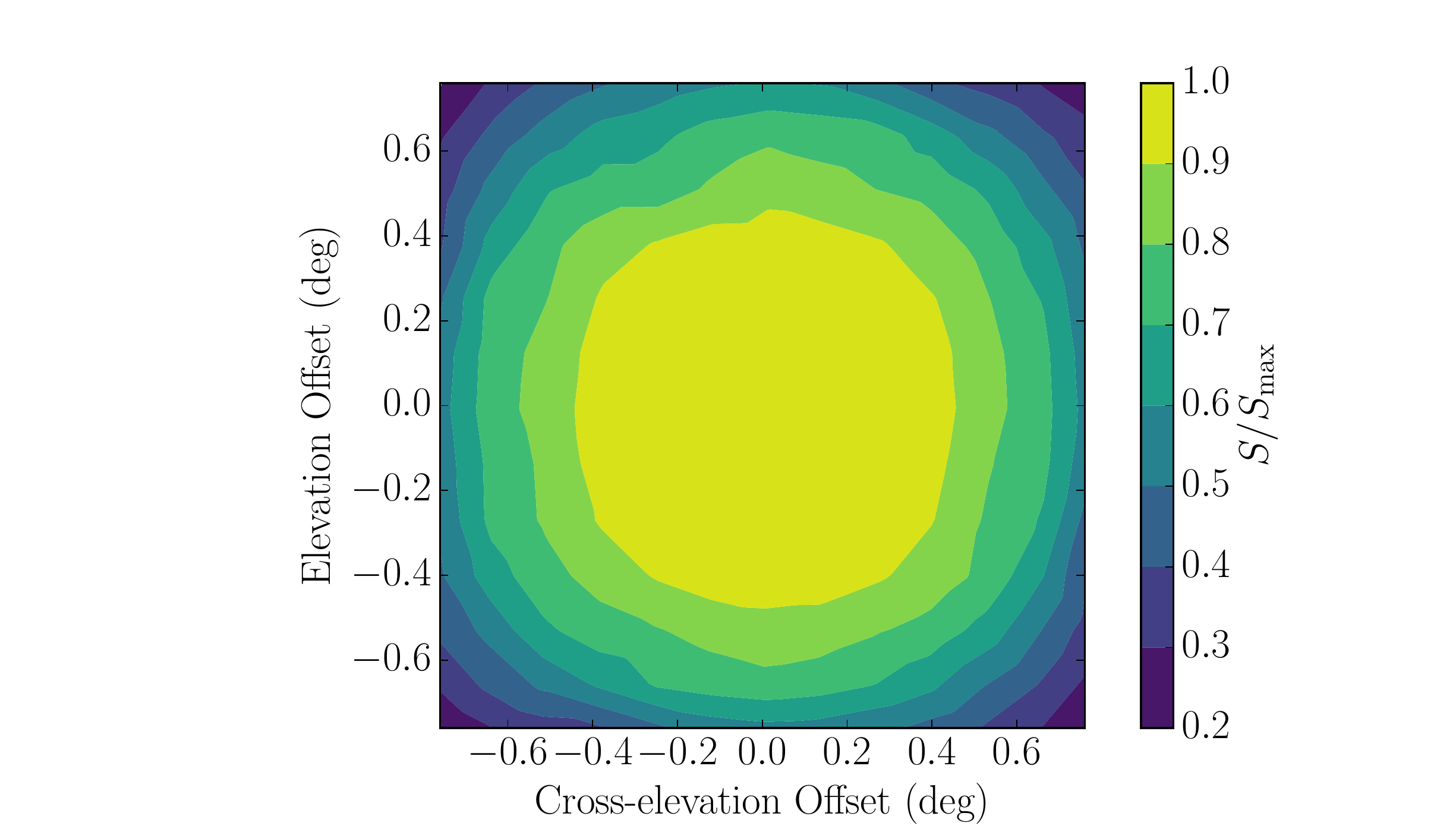}} \caption{\small Relative Stokes-I sensitivity $S/S_\text{max}$ of the MPIfR PAF on the Parkes 64~m antenna at 1310 MHz. \label{fig:fov}}
\end{figure}

Figure \ref{fig:fov} shows the envelope of Stokes-I sensitivity $S$ over the field-of-view normalised by the peak sensitivity $S_\text{max}$.  This was calculated by repeating the Y-factor measurement on Vir~A over a 13$\times$13 grid of antenna offset pointings with 0.13$^\circ$ pitch and using maxSNR weights calculated independently for each pointing.  Integrating $(S/S_\text{max})^2$ over the field of view results in an effective field of view of 1.4~deg$^2$ following the definition of \cite{bunton2010}.

\section{CONCLUSION}

We have demonstrated that the MPIfR PAF can achieve a noise temperature as low as 45~K and an aperture efficiency of 70\% to 80\% on the Parkes 64~m antenna.  It is encouraging that the modified filters for bands 2 and 3 reject sufficient RFI at Parkes for the radio-over-fibre architecture to work at the same operating power levels intended for the radio-quiet MRO.  Good sensitivity was also demonstrated in band 1, but we are yet to learn what fraction of time RFI conditions might exceed the limited headroom in band 1. 

\section*{Acknowledgments}

The Max-Plank Institute for Radio Astronomy financed the PAF discussed in this paper and its modification for a less radio-quiet site.  Installation, operation and testing at Parkes was supported by A. Dunning, Dr. D. Hayman, S. Hegarty, J. Kanapathippillai, M. Marquarding, B. Preisig, Dr. W. Raja, T. Ruckley, M. Smith, and  Dr. G. Wieching.  The Parkes radio telescope is part of the Australia Telescope National Facility which is funded by the Australian Government for operation as a National Facility managed by CSIRO.



\begin{thebibliography}{9}
\itemsep=0ex
\providecommand{\natexlab}[1]{#1}
\providecommand{\url}[1]{\texttt{#1}}
\expandafter\ifx\csname urlstyle\endcsname\relax
  \providecommand{\doi}[1]{doi: #1}\else
  \providecommand{\doi}{doi: \begingroup \urlstyle{rm}\Url}\fi


\bibitem[Thornton et al.(2013)]{Thornton2013} Thornton, D., Stappers, B., Bailes, M., et al.\ 2013, Science, 341, 53 

\bibitem[DeBoer et~al.(2009)]{DeBoer2009}
D.~R. DeBoer et~al.
\newblock Australian {SKA} Pathfinder: A high-dynamic range wide-field of view
  survey telescope.
\newblock \emph{Proc. {IEEE}}, 97\penalty0 (8):\penalty0 1507--1521, Aug. 2009.


\bibitem[Wilson et~al.(2013)Wilson, Storey, and Tzioumis]{Wilson2013}
C.~Wilson, M.~Storey, and T.~Tzioumis.
\newblock Measures for control of {EMI} and {RFI} at the {Murchison
  Radioastronomy Observatory, Australia}.
\newblock In \emph{Electromagnetic Compatibility (APEMC), 2013 Asia-Pacific
  Symposium on}, pp. 1--4, May. 2013.

\bibitem[Hampson et~al.(2012)]{Hampson2012}
G.~Hampson et~al.
\newblock {ASKAP PAF ADE} -- advancing an {L}-band {PAF} design towards {SKA}.
\newblock In \emph{Electromagnetics in Advanced Applications (ICEAA), 2012
  International Conference on}, pp. 807--809, Sept. 2012.

\bibitem[Hay and {O'Sullivan}(2008)]{Hay2008}
S.~G. Hay and J.~D. {O'Sullivan}.
\newblock Analysis of common-mode effects in a dual-polarized planar
  connected-array antenna.
\newblock \emph{Radio Sci.}, 43\penalty0 (6), Dec. 2008.


\bibitem[{Chippendale} et~al.(2016)]{Chippendale2016}
A.~P. {Chippendale} et~al.
\newblock Recent developments in measuring signal and noise in phased array
  feeds at {CSIRO}.
\newblock \emph{Antennas and Propagation (EuCAP), 2016 European Conference
  on}, Apr. 2016. 

\bibitem[Chippendale et~al.(2015{\natexlab{a}})]{Chippendale2015}
A.~P. Chippendale et~al.
\newblock Measured sensitivity of the first {M}ark {II} phased array feed on an
  {ASKAP} antenna.
\newblock In \emph{Electromagnetics in Advanced Applications (ICEAA), 2015
  International Conference on}, Nov. 2015{\natexlab{a}}.

\bibitem[Ott et~al.(1994)]{Ott1994}
M.~Ott et~al.
\newblock {An updated list of radio flux density calibrators}.
\newblock \emph{Astronomy and Astrophysics}, 284:\penalty0 331--339, Apr. 1994.

\bibitem[Chippendale et~al.(2015{\natexlab{b}})]{Chippendale2015a}
A.~P. Chippendale et~al.
\newblock Measured aperture-array noise temperature of the {M}ark {II} phased
  array feed for {ASKAP}.
\newblock In \emph{Antennas and Propagation ({ISAP}), 2015 International
  Symposium on}, Sept. 2015{\natexlab{b}}.

\bibitem[Bunton and Hay(2010)]{bunton2010}
J.~D. Bunton and S.~G. Hay.
\newblock Achievable field of view of chequerboard phased array feed.
\newblock In \emph{Electromagnetics in Advanced Applications (ICEAA), 2010
  International Conference on}, pp. 728--730, Sept. 2010.
\end{thebibliography}


\end{document}